\documentclass[reprint,amsmath,amssymb,prl]{revtex4-1}
\usepackage{graphicx,color}
\usepackage{dcolumn}
\usepackage{bm}
\usepackage{soulutf8}

\definecolor{nred}{RGB}{224,0,0}
\definecolor{nblue}  {RGB}{28,130,185}
\definecolor{dgreen} {RGB}{78,138,21}
\definecolor{norange}{RGB}{230,120,20}

\begin{document}


\title{Conformal invariance in the nonperturbative renormalization group: a rationale for choosing the regulator}

\author{Ivan Balog}
\affiliation{Institute of Physics, Bijeni\v{c}ka cesta 46, HR-10001 Zagreb, Croatia}

\author{Gonzalo De Polsi}
\affiliation{Instituto de F\'isica, Facultad de Ciencias, Universidad de la
		Rep\'ublica, Igu\'a 4225, 11400, Montevideo, Uruguay}

\author{Matthieu Tissier}
\affiliation{Sorbonne Universit\'e, CNRS, Laboratoire de Physique Th\'eorique de la Mati\`ere Condens\'ee, LPTMC, F-75005 Paris, France}

\author{Nicol\'as Wschebor}
\affiliation{Instituto de F\'isica, Facultad de Ingenier\'ia, Universidad de la
		Rep\'ublica, J.H.y Reissig 565, 11000 Montevideo, Uruguay}

\date{\today}

\begin{abstract}
Field-theoretical calculations performed in an approximation scheme often present a spurious dependence of physical quantities on some unphysical parameters associated with the details of the calculation setup (such as, the renormalization scheme or, in perturbation theory, the resummation procedure).  In the present article, we propose to reduce this dependence by invoking conformal invariance. Using as a benchmark the three-dimensional Ising model, we show that, within the derivative expansion at order 4, performed in the nonperturbative renormalization group formalism, the identity associated with this symmetry is not exactly satisfied. The calculations which best satisfy this identity are shown to yield  critical  exponents  which  coincide  to  a  high  accuracy  with  those  obtained  by  the  conformal bootstrap.
\end{abstract}

\maketitle


{\it Introduction.} In the seventies, Polyakov and Migdal conjectured that scale invariance induces conformal
invariance in a large class of systems \cite{Polyakov:1970xd,Migdal:1971xh}. Studying scale invariance via Renormalization Group (RG)
has since developed into an extensive and general toolset which enabled quantitative description
of a plethora of physical systems, with and without disorder, at- and out-of-equilibrium, with short or long range
interactions, etc... 
Unfortunately, no similar construction applies to the case of conformal invariance. Such an approach would give more constraints on the theory because special conformal invariance implies that there are $d$ more generators of symmetry in a $d$-dimensional system with $d>2$.
A natural question therefore arises: how can we implement conformal invariance within the renormalization group framework?
In this letter, we address this question in the framework of the Nonperturbative Renormalization group (NPRG)
\footnote{Similar ideas could be applied to, say, the Polchinski approach}. This modern version of Wilson's renormalization group was
developed more than two decades ago for addressing different problems in field
theories \cite{Wetterich:1992yh,Ellwanger:1993kk,Morris:1993qb,Berges:2000ew,Delamotte:2007pf}. It can be applied to many physical situations and it is nowadays widely used in solid state physics,
statistical mechanics (both in- and out-of-equilibrium), particle physics \cite{Gies:2006wv}, quantum gravity \cite{Reuter:2019byg}, etc. 
Just to give statistical mechanical examples, this method permits to access different observables, including universal
(critical exponents, amplitude ratios, scaling functions, etc.)
\cite{Seide99,VonGersdorff:2000kp,Canet:2003qd,Tarjus:2004wyx,Benitez09,Canet10,Tissier:2011zz,Benitez:2011xx,Canet11,Leonard15,Rose:2016wqz,Balog:2019rrg,DePolsi:2020pjk}
but also non-universal quantities (phase diagrams, critical temperatures, etc.) \cite{Parola95,Seide99,Canet:2003yu,Canet:2004je,Caillol06,Machado:2010wi,Rancon12}. 

In a nutshell, NPRG consists in regularizing a theory, defined by an action (or hamiltonian) $S$, by adding to $S$ a quadratic scale-dependent term $\Delta S_k=1/2\int_q R_k(q) \chi(q) \chi(-q)$. The regulating function $R_k(q)$ separates the high and low momentum modes of $\chi(q)$ by freezing the latter while leaving unchanged the former. It is then possible to deduce an exact flow equation
\cite{Wetterich:1992yh,Ellwanger:1993kk,Morris:1993qb}, usually denoted the Wetterich equation, 
which governs the evolution of a
scale-dependent effective action $\Gamma_k$ (which closely resembles the Gibbs free energy) when the RG scale $k$ is reduced. The exact flow equation is a
nonlinear functional equation which cannot be solved exactly. In actual calculations, the general strategy consists in looking
for {\it approximate} solutions to this {\it exact} equation: One considers a family of effective actions, typically parametrized by
a few functions, and solves the exact equation within this functional subspace of actions. The strength of the method is that it is resilient
to approximations: many qualitative features of the theory can be reproduced with quite crude approximations. In more quantitative studies,
a systematic approximation scheme, called derivative expansion (DE) proved to be very efficient \cite{Berges:2000ew,Delamotte:2007pf}.
The order $\mathcal{O}(\partial^s)$ of the approximation, consists in keeping all terms with up to $s$ gradients in a gradient expansion of the effective action $\Gamma_k[\phi]$. Recent studies showed that this enables one to determine
the critical exponents of the Ising and O($N$) model to a high precision if enough terms are retained in the
approximation \cite{Canet:2003qd,Balog:2019rrg,DePolsi:2020pjk}. To achieve this level of precision, a major drawback must be overcome.
Within the
DE, the values of physical quantities depend on the regulator $R_k(q)$ which is used. This spurious dependence is a consequence of the
approximation as no such dependence should show up in the exact theory at $k=0$. Therefore, the determination of physical quantities such as the
critical exponents relies on choosing some sort of an optimal regulator. By lack of a better choice,
the common approach consists in invoking the Principle of Minimal Sensitivity (PMS) \cite{Stevenson:1981vj}: the optimal regulator is chosen
such that the critical
exponents are locally independent of the regulator \cite{Canet:2002gs,Balog:2019rrg,DePolsi:2020pjk}. Once critical exponents are studied in the
light of the PMS, the DE proved to be a very powerful method yielding very accurate critical exponents in the Ising and O($N$) universality
classes \cite{Balog:2019rrg,DePolsi:2020pjk}. The use of the PMS has been criticized because we may encounter situations where no local
extremum exists, or where many extrema are present. In these situations, it is not clear which regulator should be retained. We stress
that a similar spurious dependence on parameters is observed in other approaches. For instance, in the framework of perturbation theory,
series are only asymptotic and it is necessary to resort to resummation techniques to compute
critical exponents. These always come with some parameters that must be fixed in one way or
another, see {\it e.g.} \cite{Guida:1998bx,Kompaniets:2017yct}. 

Our aim in this letter is to combine NPRG techniques and the fact that critical physics is conformal invariant. The relation between these
two concepts was already discussed in Ref.~\cite{delamotte2016scale,Rosten:2016zap,DePolsi2019} where it was shown that if the theory
is invariant under the full conformal group, the effective action satisfies, on top of the fixed point equation which encodes scale
invariance, a very similar constraint which accounts for  special conformal transformations.
Our strategy then consists in studying this constraint within the DE.  We focus on the simplest situation where
conformal invariance gives more information than the sole  scale invariance: the Ising universality class at order $\mathcal O(\partial^4)$.
We check explicitly that the special conformal constraint is not exactly fulfilled at the fixed point. This is an artifact of our approximation
procedure because, in the exact theory, the fixed-point solution should fulfill conformal invariance \cite{delamotte2016scale}.
We can turn this drawback into an asset as it calls for choosing the regulator which yields the smallest breaking of conformal invariance.
We propose to call this the Principle of Maximal Conformality (PMC). Explicit calculations show that PMC leads to critical exponents
which are very close to the almost exact conformal bootstrap results. In fact, PMC and PMS lead to very similar regulators.
This gives strong indication that
the PMS procedure is reasonable, in the framework of NPRG and gives us more confidence on the critical exponents
obtained in Refs.~\cite{Canet:2002gs,Canet:2003qd,Balog:2019rrg,DePolsi:2020pjk}.

{\it NPRG and conformal invariance.} The dynamics of the model is characterized by a microscopic action $S$ which we choose invariant under translations, rotations and
under $\mathbb Z_2$ symmetry. Following Wetterich \cite{Wetterich:1992yh}, we introduce a regularized partition function $Z_R$  as:
\begin{align}
\label{eq_Z}
    Z_R[J]&=\int \mathcal D\chi e^{-S[\chi]+\int_x J(x)\chi(x)-\frac 12 \int_{xy}\chi(x) R(x,y)\chi(y)}.
\end{align}
The regulating function $R(x,y)$ is in general chosen to be invariant under translations. Its Fourier transform $R_k(q)$ is of order $k^2$ at small momentum and tends to zero at large $q$. This ensures that the long-distance modes are frozen. 
As usual, the order parameter is obtained as a functional derivative of $W_R=\log Z_R$ with respect to $J$
\begin{equation}
    \phi(x)=\frac{\delta W_R[J]}{\delta J(x)}.
\end{equation}
It is convenient to define a scale-dependent effective  action $\Gamma_R$ as the Legendre transform of $W_R$ with respect to the source $J$, where a part quadratic in the field is subtracted:
\begin{equation}
    \Gamma_R[\phi]+W_R[J]=\int_xJ(x)\phi(x)-\frac 12\int_{xy}\phi(x)R(x,y)\phi(y).
\end{equation}

Now consider a particular family of regulators $R_k$, indexed by a momentum scale $k$ and renormalized fields $\tilde \phi$ such that:
\begin{align}
    R_k(x,y)&=k^{D_R} \tilde R(k^2(x-y)^2),\\
    \phi(x)&=k^{D_\phi}\tilde\phi(k x),
\end{align}
where $D_\phi$ is the dimension of the field (to be fixed later on) and $D_R=2(d-D_\phi)$. An exact flow equation
can be obtained by deriving the effective action with respect to $k$ (or, equivalently, with respect to the
renormalization-group ``time'' $t=\log k/\Lambda$ where $\Lambda$ is an arbitrary reference scale) at
fixed renormalized field $\tilde \phi$ and renormalized regulator $\tilde R$
\begin{equation}
\label{eq_NPRG}
    \partial_t\Gamma_{R_k}|_{\tilde \phi,\tilde R}=(\mathcal D_\phi+\mathcal D_R)\Gamma_{R_k},
\end{equation}
where we introduced the operators
\begin{align}
\label{eq_df}
    \mathcal D_\phi&=\int_x [(D_\phi +x^\mu\partial_{x^\mu}) \phi(x)]\frac{\delta\quad }{\delta\phi(x)}\\
    \label{eq_dR}    \mathcal D_R&=\int_{xy} [(D_R +x^\mu\partial_{x^\mu}+y^\mu\partial_{y^\mu} )R(x,y)]\frac{\delta\qquad }{\delta R(x,y)}
\end{align}
The important observation at this point is that the functional derivative of $\Gamma_R$ with respect to 
$R$ can be expressed in terms of the exact propagator of the theory:
\begin{equation}
\label{eq_prop}
    \frac{\delta \Gamma_R}{\delta R(x,y)}=\frac 12\left( \Gamma_R^{(2)}(x,y)+R(x,y)\right)^{-1}
\end{equation}
In this way, we obtain an exact flow equation which involves only functional derivatives with respect to the field, not with respect to the regulating function. 

The flow equation (\ref{eq_NPRG}) makes it clear that at a fixed point ($\partial_t \Gamma_{R_k}=0$), the effective action is invariant
under a simultaneous dilatation of the field and of the regulator, according to the laws given in Eqs.~(\ref{eq_df},\ref{eq_dR}). A similar
constraint can be derived for a field theory invariant under conformal transformations.
As shown in Ref.~\cite{delamotte2016scale,Rosten:2016zap,DePolsi2019}, if a theory is conformal invariant, when regularized as in  Eq.~(\ref{eq_Z}), its effective action $\Gamma_{R}$ is invariant under a simultaneous change of the field and regulator:
\begin{equation}
\label{eq_conf}
    0=(\mathcal K^\mu_\phi+\mathcal K^\mu_R)\Gamma_R,
\end{equation}
where we introduced the operators
\begin{align}
\label{eq_kf}
    \mathcal K^\mu_\phi=\int_x& [( K^\mu_x+2x^\mu D_\phi) \phi(x)]\frac{\delta\quad }{\delta\phi(x)},\\
    \mathcal K^\mu_R=\int_{xy}& [K^\mu_x+K^\mu_y+ D_R(x^\mu+y^\mu) )R(x,y)]\frac{\delta\qquad }{\delta R(x,y)},\nonumber
\end{align}
and $K^\mu_x=2 x^\mu x^\nu\partial_{x^\nu}-x^2 \partial_{x^\mu}$.
Equations (\ref{eq_conf}) and  (\ref{eq_NPRG}) at the fixed point, supplemented by Eq. (\ref{eq_prop}), are the Ward identities associated with special conformal and dilatation invariance respectively, in the presence of a regulator. These Ward identities are functional constraints  that are difficult to handle. It proves more convenient in practice to compute the functional derivative of these equations with respect to $\phi(z_a)\cdots \phi(z_n)$, evaluate them in a homogeneous field configuration $\phi(x)=\phi$ and take the Fourier transform with respect to the space coordinates. In this last step, we use translational invariance to put one coordinate of the $n$-point vertex at the origin and Fourier-transform with respect to the $n-1$ other coordinates. We thus get an infinite tower of coupled equations which involve $n$-point vertex functions:
\begin{align}
        &\left(\sum_{i=1}^{n-1}p_i^\mu\partial_{p_i^\mu}+n D_\phi-d +\phi D_\phi\partial_\phi\right)\Gamma_{R_k}^{(n)}(P) \nonumber \\
        &=\int_q \dot R(q)H^{(n)}(q,-q;P),
\label{eq_dilatn}
\end{align}
\begin{align}
&\sum_{i=1}^{n-1}\left[\tilde K_{p_i}^\mu
    +2 D_\phi\partial_{p_i^\mu}\right]\Gamma_{R_k}^{(n)}(P)+2\phi D_\phi\partial_{r^\mu}^\star \Gamma_{R_k}^{(n+1)}(r,P)\nonumber \\
    & =\int_q \dot R(q)\partial_{r^\mu}^\star H^{(n)}(r+q,r-q;P),\label{eq_confn}
\end{align}
where $\tilde K^\mu_p=2p^\nu\partial_{p^\mu}\partial_{p^\nu}-p^\mu\partial_{p^\nu}^2$, $H^{(n)}$ is the Fourier transform
of $-(\delta^{n+1}\Gamma_{R})/(\delta R(x,y)\delta \phi(z_1)\cdots\delta \phi(z_n))\Big|_{R=R_k}$, \footnote{By using Eq.~(\ref{eq_prop}), $H^{(n)}$ can be
expressed in terms of propagators and vertex functions.} the first two momenta being conjugate to $x$ and $y$, $\dot R(q)$ is the Fourier
transform of $(D_R +x^\mu\partial_{x^\mu}+y^\mu\partial_{y^\mu} )R(x,y)$, for a regulating function which is invariant under translations
and rotations, $P$ represents the $n-1$ independent external momenta, the star in $\partial_{r^\mu}^\star$ indicates that the momentum $r$
must be set to zero after the derivative is performed. 

As a direct consequence of its definition, the function $H$ is symmetric under permutations of the external momenta $p_i$ so that both Eqs.~(\ref{eq_dilatn}) and (\ref{eq_confn}) share this property. The former is also invariant under replacing $p_1$ by $-\sum_{i=1}^{n-1}p_i$, a property which is not shared by the latter. These equations are therefore S$_n$ and S$_{n-1}$ invariant respectively. It is convenient to restore the S$_n$ in the conformal Ward identity. This can be done by applying the operator  $-2/n(\sum_{i=1}^{n-1}\partial_{p_i^\mu})$ to the Ward identity for dilatation (\ref{eq_dilatn}) and adding to this the conformal Ward identity (\ref{eq_confn}). From then on, we make use only of this symmetrized version of the Ward identity for special conformal invariance.

{\it Derivative expansion.} Let us now analyze the consequences of conformal invariance within the derivative expansion. At order $\mathcal O(\partial^4)$, on which
we concentrate in this article, we have to introduce five functions of the homogeneous field $\phi$ and the truncation reads:
\begin{equation}
\label{eq_ansatz}
\begin{split}
    \Gamma_{R_k}[\phi]&=\int_x\Big\{U_k+\frac 12 Z_k(\partial_\mu\phi)^2+\frac 12 W_{a,k}(\partial_\mu\partial_\nu\phi)^2\\&+\frac \phi2 W_{b,k}(\partial^2\phi)(\partial_\nu\phi)^2+\frac 12W_{c,k}[(\partial_\mu\phi)^2]^2\Big\}.
\end{split}
\end{equation}
Because of the $\mathbb Z_2$ symmetry, $U_k(\phi)$, $Z_k(\phi)$, $W_{a,k}(\phi)$, $W_{b,k}(\phi)$ and $W_{c,k}(\phi)$ are even
functions of $\phi$. From now on, all functions are expressed in terms of renormalized and dimensionless field $\tilde{\varphi}$ that are functions of the dimensionless coordinates $\tilde{x}=kx$.

It is interesting, at this point to count how many independent structures in momenta appear in the symmetrized version of Eq.~(\ref{eq_conf}).
This can be entirely determined by using the S$_n$ symmetry, the conservation of momentum and the fact that Eq.~(\ref{eq_conf}) has one vector
index $\mu$. We find by inspection that there is no such tensorial structure of order $\mathcal O(p^1)$ (the only S$_n$-symmetric
combination of $n$ vectors is the sum of these vectors, which vanishes by conservation of the momentum).
There is just one structure at order $\mathcal O(p^3)$: $\sum_i p_i^2 p_i^\mu$\footnote{There are 4 structures at order $\mathcal O( p^5)$ and
15 at order $\mathcal O( p^7)$.}. Observe that this invariant exists only for structures with at least 3 external momenta.
This is why special conformal invariance starts to play a role at order $\mathcal O(\partial^4)$ in the DE.
 
As a consequence, the DE at order $\mathcal{O}(\partial^4)$ is characterized by 5 functions of $\tilde{\phi}$. At the fixed point controlling the critical regime, these functions are completely determined by scale invariance, Eq.~(\ref{eq_dilatn}). Moreover, there is one further constraint that emerges from the symmetrized Ward identity for conformal invariance which makes the problem over-constrained.
All 6 functional constraints involve a ``scaling contribution'' and a ``loop contribution'' which originate from the left-hand-sides and
right-hand-sides  of Eqs.~(\ref{eq_dilatn},\ref{eq_confn}) respectively. The loop contributions are obtained by computing the right-hand-sides of (\ref{eq_dilatn},\ref{eq_confn}) with the explicit form of the truncation and projecting the resulting expression on the same functional space. Of course, the right-hand-sides also involve terms with higher powers of the external momenta and the approximation consists in neglecting these terms. In what concerns scale invariance,
these equations were computed in \cite{Canet:2003qd,delamotte2016scale,DePolsi2019} and we use the same expressions
in the present work 
\footnote{We employed the equations of Refs.~\cite{delamotte2016scale,DePolsi2019} that differ with
those of Ref.~\cite{Canet:2003qd} by terms of order $\mathcal O(\partial^6)$.}.
We, moreover, treat the symmetrized version of the conformal constraint (\ref{eq_conf}) at order $\mathcal O(p^3)$ in the same manner.
All these expressions are rather lengthy and are presented in the Supplemental Material. 

\begin{figure}[t]
    \centering
    \includegraphics[width=.8\linewidth]{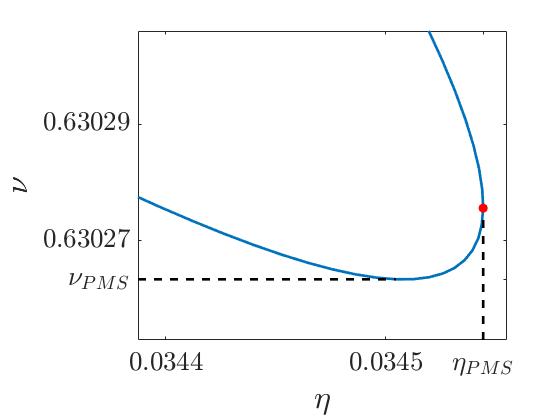}
    \caption{Critical exponents obtained for different values of $\alpha$. The PMS predictions correspond to the extremal values of these
    exponents, indicated by dashed lines. The value of $\alpha$ which minimizes $f(0)$ leads to exponents indicated by a point.}
    \label{fig_exp}
\end{figure}

To quantify the breaking of the conformal constraint, we consider the constraint that appears at order $\mathcal{O}(p^3)$ normalized by its ``scaling" contribution at $\phi=0$. This defines the function $f(\phi)$ (see the Supplemental Material for the expressions for the conformal constraint as well as the function $f$), which would be exactly zero if conformal invariance was realized within DE.
\begin{figure}[tbp]
    \centering
    \includegraphics[width=\linewidth]{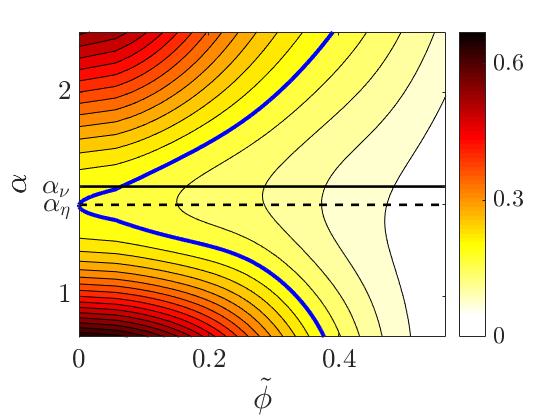}
    \caption{Contour plot of the function $f$ (see the text) as a function of $\tilde \phi$ (vertical axis) for various values of $\alpha$ (horizontal axis). The blue level curve corresponds to the minimal value of $f(\tilde \phi=0)$.  The solid and dashed lines correspond to the values of $\alpha$ which fulfill the PMS criterion for the exponent $\nu$ and $\eta$ respectively.}
    \label{fig_error}
\end{figure}
We now study how the breaking of conformal invariance depends on the regulator. We focus the discussion on the
regulator $R_k=\alpha q^2/[\exp(q^2/k^2)-1]$ which has been widely
employed in previous analysis \cite{Wetterich:1992yh}, but we have checked that our findings are not modified if we use any of the regulator families considered in Refs.~\cite{delamotte2016scale,DePolsi2019}.
We plot in Fig.~\ref{fig_error} the function $f(\tilde\phi)$ for various values of $\alpha$. We observe that $f$ is always strictly positive.
It is thus convenient to choose the regulator according to the Principle of Maximal Conformality, such that the breaking
of conformal invariance is as small as possible. To do so,  observe the first striking feature depicted in this plot: for each value of $\tilde\phi$, $f$ reaches a minimum for very similar values of $\alpha$. As a consequence, the different
implementations of the PMC (i.e. minimizing $f(0)$ or the integral of $f(\tilde\phi)$ with some weight, etc.) lead to very similar values of $\alpha$ (see Fig.~\ref{fig_error}). In what follows, we choose to minimize $f(0)$ and call $\alpha_c$ the associated value of $\alpha$. This value of $\alpha$ corresponds to the intersection of the blue curve with the vertical axis in Fig.~\ref{fig_error}. The other striking feature of Fig.~\ref{fig_error} is that $\alpha_c$ is very close to the optimal values of $\alpha$ obtained by using the PMS criterion. Thus, the critical exponents obtained by the PMC are very close to those obtained by using the PMS. This is depicted in Fig.~\ref{fig_exp} where we present a
parametric plot of the critical exponents $\eta$ and $\nu$ for different values of $\alpha$. The PMS critical exponents are easily spotted
as the extremal points. To be more quantitative, we obtain with the PMC $\alpha_c=1.45(1)$ while the PMS for the exponents $\eta$ and $\nu$
yield $\alpha_\eta=1.45(1)$ and $\alpha_\nu=1.54(1)$. Our determination of the critical exponents are depicted in Fig.~\ref{fig_exp}.
At the PMC point, we find $\eta=0.034544$ and $\nu=0.630276$ which compare well with the PMS results, $\eta=0.034544$, $\nu=0.630263$ and the
conformal bootstrap results $\eta=0.036298(2)$ and
$\nu=0.629971(4)$ \cite{Kos:2014bka}.

{\it Conclusions.} We have studied how much the Ward identity for conformal invariance is violated at the fixed point describing the $d=3$ Ising universality class,
obtained within the derivative expansion at order $\mathcal O(\partial^4)$. Since this symmetry is known to be realized at the critical point of
this model, the violation of the associated Ward identity is an unfortunate consequence of the approximation that was performed.
In this letter, we propose to choose the regulator according to the Principle of Maximal Conformality: We have considered
a 1-parameter family of regulators, characterized by a multiplicative constant $\alpha$ and have searched for the value of $\alpha$ which yields
the smallest breaking of the Ward identity for conformal invariance. This leads to very precise determinations of the critical exponents,
which differ from those obtained by conformal bootstrap only by a small fraction. The optimal value of $\alpha$
is also surprisingly
close to the one obtained by the Principle of Minimal Sensitivity which is empirically known to yield quite precise determinations of
critical exponents.
This study clearly demonstrates the power of the Principle of Maximal Conformality, which could be used in other models within the NPRG method.
It also supports the use of the Principle of Minimal Sensitivity in the framework of the NPRG. Generalization to other approaches, such
as resummation methods in perturbation theory could also be considered. 

{\it Acknowledgments.} The authors thank M. V. Kompanietz, E. Panzer and G. Tarjus for useful discussions. The authors would also like to thank B. Delamotte for useful discussions and a very careful reading of a previous version of this manuscript. This work was supported by Grant I+D number 412 of the
CSIC (UdelaR) Commission and Programa de Desarrollo de las Ciencias B\'asicas (PEDECIBA), Uruguay
and ECOS Sud U17E01. IB acknowledges the support of the Croatian Science Foundation Project IP-2016-6-7258
and the QuantiXLie Centre of Excellence,
a project cofinanced by the Croatian Government and
European Union through the European Regional Development Fund - the Competitiveness and Cohesion Operational
Programme (Grant KK.01.1.1.01.0004).

\bibliographystyle{apsrev4-1} 
\bibliography{Conformal_nprg}

\begin{thebibliography}{40}%
\makeatletter
\providecommand \@ifxundefined [1]{%
 \@ifx{#1\undefined}
}%
\providecommand \@ifnum [1]{%
 \ifnum #1\expandafter \@firstoftwo
 \else \expandafter \@secondoftwo
 \fi
}%
\providecommand \@ifx [1]{%
 \ifx #1\expandafter \@firstoftwo
 \else \expandafter \@secondoftwo
 \fi
}%
\providecommand \natexlab [1]{#1}%
\providecommand \enquote  [1]{``#1''}%
\providecommand \bibnamefont  [1]{#1}%
\providecommand \bibfnamefont [1]{#1}%
\providecommand \citenamefont [1]{#1}%
\providecommand \href@noop [0]{\@secondoftwo}%
\providecommand \href [0]{\begingroup \@sanitize@url \@href}%
\providecommand \@href[1]{\@@startlink{#1}\@@href}%
\providecommand \@@href[1]{\endgroup#1\@@endlink}%
\providecommand \@sanitize@url [0]{\catcode `\\12\catcode `\$12\catcode
  `\&12\catcode `\#12\catcode `\^12\catcode `\_12\catcode `\%12\relax}%
\providecommand \@@startlink[1]{}%
\providecommand \@@endlink[0]{}%
\providecommand \url  [0]{\begingroup\@sanitize@url \@url }%
\providecommand \@url [1]{\endgroup\@href {#1}{\urlprefix }}%
\providecommand \urlprefix  [0]{URL }%
\providecommand \Eprint [0]{\href }%
\providecommand \doibase [0]{http://dx.doi.org/}%
\providecommand \selectlanguage [0]{\@gobble}%
\providecommand \bibinfo  [0]{\@secondoftwo}%
\providecommand \bibfield  [0]{\@secondoftwo}%
\providecommand \translation [1]{[#1]}%
\providecommand \BibitemOpen [0]{}%
\providecommand \bibitemStop [0]{}%
\providecommand \bibitemNoStop [0]{.\EOS\space}%
\providecommand \EOS [0]{\spacefactor3000\relax}%
\providecommand \BibitemShut  [1]{\csname bibitem#1\endcsname}%
\let\auto@bib@innerbib\@empty
\bibitem [{\citenamefont {Polyakov}(1970)}]{Polyakov:1970xd}%
  \BibitemOpen
  \bibfield  {author} {\bibinfo {author} {\bibfnamefont {A.~M.}\ \bibnamefont
  {Polyakov}},\ }\href@noop {} {\bibfield  {journal} {\bibinfo  {journal} {JETP
  Lett.}\ }\textbf {\bibinfo {volume} {12}},\ \bibinfo {pages} {381} (\bibinfo
  {year} {1970})},\ \bibinfo {note} {[Pisma Zh. Eksp. Teor.
  Fiz.12,538(1970)]}\BibitemShut {NoStop}%
\bibitem [{\citenamefont {Migdal}(1971)}]{Migdal:1971xh}%
  \BibitemOpen
  \bibfield  {author} {\bibinfo {author} {\bibfnamefont {A.~A.}\ \bibnamefont
  {Migdal}},\ }\href {\doibase 10.1016/0370-2693(71)90583-1} {\bibfield
  {journal} {\bibinfo  {journal} {Phys. Lett.}\ }\textbf {\bibinfo {volume}
  {37B}},\ \bibinfo {pages} {98} (\bibinfo {year} {1971})}\BibitemShut
  {NoStop}%
\bibitem [{Note1()}]{Note1}%
  \BibitemOpen
  \bibinfo {note} {Similar ideas could be applied to, say, the Polchinski
  approach}\BibitemShut {NoStop}%
\bibitem [{\citenamefont {Wetterich}(1993)}]{Wetterich:1992yh}%
  \BibitemOpen
  \bibfield  {author} {\bibinfo {author} {\bibfnamefont {C.}~\bibnamefont
  {Wetterich}},\ }\href {\doibase 10.1016/0370-2693(93)90726-X} {\bibfield
  {journal} {\bibinfo  {journal} {Phys. Lett.}\ }\textbf {\bibinfo {volume}
  {B301}},\ \bibinfo {pages} {90} (\bibinfo {year} {1993})},\ \Eprint
  {http://arxiv.org/abs/1710.05815} {arXiv:1710.05815 [hep-th]} \BibitemShut
  {NoStop}%
\bibitem [{\citenamefont {Ellwanger}(1993)}]{Ellwanger:1993kk}%
  \BibitemOpen
  \bibfield  {author} {\bibinfo {author} {\bibfnamefont {U.}~\bibnamefont
  {Ellwanger}},\ }\href {\doibase 10.1007/BF01553022} {\bibfield  {journal}
  {\bibinfo  {journal} {Z. Phys.}\ }\textbf {\bibinfo {volume} {C58}},\
  \bibinfo {pages} {619} (\bibinfo {year} {1993})}\BibitemShut {NoStop}%
\bibitem [{\citenamefont {Morris}(1994)}]{Morris:1993qb}%
  \BibitemOpen
  \bibfield  {author} {\bibinfo {author} {\bibfnamefont {T.~R.}\ \bibnamefont
  {Morris}},\ }\href {\doibase 10.1142/S0217751X94000972} {\bibfield  {journal}
  {\bibinfo  {journal} {Int. J. Mod. Phys.}\ }\textbf {\bibinfo {volume}
  {A9}},\ \bibinfo {pages} {2411} (\bibinfo {year} {1994})},\ \Eprint
  {http://arxiv.org/abs/hep-ph/9308265} {arXiv:hep-ph/9308265 [hep-ph]}
  \BibitemShut {NoStop}%
\bibitem [{\citenamefont {Berges}\ \emph {et~al.}(2002)\citenamefont {Berges},
  \citenamefont {Tetradis},\ and\ \citenamefont {Wetterich}}]{Berges:2000ew}%
  \BibitemOpen
  \bibfield  {author} {\bibinfo {author} {\bibfnamefont {J.}~\bibnamefont
  {Berges}}, \bibinfo {author} {\bibfnamefont {N.}~\bibnamefont {Tetradis}}, \
  and\ \bibinfo {author} {\bibfnamefont {C.}~\bibnamefont {Wetterich}},\ }\href
  {\doibase 10.1016/S0370-1573(01)00098-9} {\bibfield  {journal} {\bibinfo
  {journal} {Phys. Rept.}\ }\textbf {\bibinfo {volume} {363}},\ \bibinfo
  {pages} {223} (\bibinfo {year} {2002})},\ \Eprint
  {http://arxiv.org/abs/hep-ph/0005122} {arXiv:hep-ph/0005122 [hep-ph]}
  \BibitemShut {NoStop}%
\bibitem [{\citenamefont {Delamotte}(2012)}]{Delamotte:2007pf}%
  \BibitemOpen
  \bibfield  {author} {\bibinfo {author} {\bibfnamefont {B.}~\bibnamefont
  {Delamotte}},\ }\href {\doibase 10.1007/978-3-642-27320-9_2} {\bibfield
  {journal} {\bibinfo  {journal} {Lect. Notes Phys.}\ }\textbf {\bibinfo
  {volume} {852}},\ \bibinfo {pages} {49} (\bibinfo {year} {2012})},\ \Eprint
  {http://arxiv.org/abs/cond-mat/0702365} {arXiv:cond-mat/0702365
  [cond-mat.stat-mech]} \BibitemShut {NoStop}%
\bibitem [{\citenamefont {Gies}(2012)}]{Gies:2006wv}%
  \BibitemOpen
  \bibfield  {author} {\bibinfo {author} {\bibfnamefont {H.}~\bibnamefont
  {Gies}},\ }\href {\doibase 10.1007/978-3-642-27320-9_6} {\emph {\bibinfo
  {title} {Renormalization group and effective field theory approaches to
  many-body systems}}},\ Vol.\ \bibinfo {volume} {852}\ (\bibinfo {year}
  {2012})\ Chap.\ \bibinfo {chapter} {Introduction to the functional RG and
  applications to gauge theories}, pp.\ \bibinfo {pages} {287--348},\ \Eprint
  {http://arxiv.org/abs/hep-ph/0611146} {arXiv:hep-ph/0611146 [hep-ph]}
  \BibitemShut {NoStop}%
\bibitem [{\citenamefont {Reuter}\ and\ \citenamefont
  {Saueressig}(2019)}]{Reuter:2019byg}%
  \BibitemOpen
  \bibfield  {author} {\bibinfo {author} {\bibfnamefont {M.}~\bibnamefont
  {Reuter}}\ and\ \bibinfo {author} {\bibfnamefont {F.}~\bibnamefont
  {Saueressig}},\ }\href
  {https://www.cambridge.org/academic/subjects/physics/theoretical-physics-and-mathematical-physics/quantum-gravity-and-functional-renormalization-group-road-towards-asymptotic-safety?format=HB&isbn=9781107107328}
  {\emph {\bibinfo {title} {Quantum Gravity and the Functional Renormalization
  Group}}}\ (\bibinfo  {publisher} {Cambridge University Press},\ \bibinfo
  {year} {2019})\BibitemShut {NoStop}%
\bibitem [{\citenamefont {Seide}\ and\ \citenamefont
  {Wetterich}(1999)}]{Seide99}%
  \BibitemOpen
  \bibfield  {author} {\bibinfo {author} {\bibfnamefont {S.}~\bibnamefont
  {Seide}}\ and\ \bibinfo {author} {\bibfnamefont {C.}~\bibnamefont
  {Wetterich}},\ }\href {\doibase
  http://dx.doi.org/10.1016/S0550-3213(99)00545-3} {\bibfield  {journal}
  {\bibinfo  {journal} {Nucl. Phys. B}\ }\textbf {\bibinfo {volume} {562}},\
  \bibinfo {pages} {524} (\bibinfo {year} {1999})}\BibitemShut {NoStop}%
\bibitem [{\citenamefont {Von~Gersdorff}\ and\ \citenamefont
  {Wetterich}(2001)}]{VonGersdorff:2000kp}%
  \BibitemOpen
  \bibfield  {author} {\bibinfo {author} {\bibfnamefont {G.}~\bibnamefont
  {Von~Gersdorff}}\ and\ \bibinfo {author} {\bibfnamefont {C.}~\bibnamefont
  {Wetterich}},\ }\href {\doibase 10.1103/PhysRevB.64.054513} {\bibfield
  {journal} {\bibinfo  {journal} {Phys. Rev.}\ }\textbf {\bibinfo {volume}
  {B64}},\ \bibinfo {pages} {054513} (\bibinfo {year} {2001})},\ \Eprint
  {http://arxiv.org/abs/hep-th/0008114} {arXiv:hep-th/0008114 [hep-th]}
  \BibitemShut {NoStop}%
\bibitem [{\citenamefont {Canet}\ \emph
  {et~al.}(2003{\natexlab{a}})\citenamefont {Canet}, \citenamefont {Delamotte},
  \citenamefont {Mouhanna},\ and\ \citenamefont {Vidal}}]{Canet:2003qd}%
  \BibitemOpen
  \bibfield  {author} {\bibinfo {author} {\bibfnamefont {L.}~\bibnamefont
  {Canet}}, \bibinfo {author} {\bibfnamefont {B.}~\bibnamefont {Delamotte}},
  \bibinfo {author} {\bibfnamefont {D.}~\bibnamefont {Mouhanna}}, \ and\
  \bibinfo {author} {\bibfnamefont {J.}~\bibnamefont {Vidal}},\ }\href
  {\doibase 10.1103/PhysRevB.68.064421} {\bibfield  {journal} {\bibinfo
  {journal} {Phys. Rev.}\ }\textbf {\bibinfo {volume} {B68}},\ \bibinfo {pages}
  {064421} (\bibinfo {year} {2003}{\natexlab{a}})},\ \Eprint
  {http://arxiv.org/abs/hep-th/0302227} {arXiv:hep-th/0302227 [hep-th]}
  \BibitemShut {NoStop}%
\bibitem [{\citenamefont {Tarjus}\ and\ \citenamefont
  {Tissier}(2004)}]{Tarjus:2004wyx}%
  \BibitemOpen
  \bibfield  {author} {\bibinfo {author} {\bibfnamefont {G.}~\bibnamefont
  {Tarjus}}\ and\ \bibinfo {author} {\bibfnamefont {M.}~\bibnamefont
  {Tissier}},\ }\href {\doibase 10.1103/PhysRevLett.93.267008} {\bibfield
  {journal} {\bibinfo  {journal} {Phys. Rev. Lett.}\ }\textbf {\bibinfo
  {volume} {93}},\ \bibinfo {pages} {267008} (\bibinfo {year} {2004})},\
  \Eprint {http://arxiv.org/abs/cond-mat/0410118} {arXiv:cond-mat/0410118
  [cond-mat.dis-nn]} \BibitemShut {NoStop}%
\bibitem [{\citenamefont {Benitez}\ \emph {et~al.}(2009)\citenamefont
  {Benitez}, \citenamefont {Blaizot}, \citenamefont {Chat\'e}, \citenamefont
  {Delamotte}, \citenamefont {M\'endez-Galain},\ and\ \citenamefont
  {Wschebor}}]{Benitez09}%
  \BibitemOpen
  \bibfield  {author} {\bibinfo {author} {\bibfnamefont {F.}~\bibnamefont
  {Benitez}}, \bibinfo {author} {\bibfnamefont {J.-P.}\ \bibnamefont
  {Blaizot}}, \bibinfo {author} {\bibfnamefont {H.}~\bibnamefont {Chat\'e}},
  \bibinfo {author} {\bibfnamefont {B.}~\bibnamefont {Delamotte}}, \bibinfo
  {author} {\bibfnamefont {R.}~\bibnamefont {M\'endez-Galain}}, \ and\ \bibinfo
  {author} {\bibfnamefont {N.}~\bibnamefont {Wschebor}},\ }\href {\doibase
  10.1103/PhysRevE.80.030103} {\bibfield  {journal} {\bibinfo  {journal} {Phys.
  Rev. E}\ }\textbf {\bibinfo {volume} {80}},\ \bibinfo {pages} {030103(R)}
  (\bibinfo {year} {2009})}\BibitemShut {NoStop}%
\bibitem [{\citenamefont {Canet}\ \emph {et~al.}(2010)\citenamefont {Canet},
  \citenamefont {Chat\'e}, \citenamefont {Delamotte},\ and\ \citenamefont
  {Wschebor}}]{Canet10}%
  \BibitemOpen
  \bibfield  {author} {\bibinfo {author} {\bibfnamefont {L.}~\bibnamefont
  {Canet}}, \bibinfo {author} {\bibfnamefont {H.}~\bibnamefont {Chat\'e}},
  \bibinfo {author} {\bibfnamefont {B.}~\bibnamefont {Delamotte}}, \ and\
  \bibinfo {author} {\bibfnamefont {N.}~\bibnamefont {Wschebor}},\ }\href
  {\doibase 10.1103/PhysRevLett.104.150601} {\bibfield  {journal} {\bibinfo
  {journal} {Phys. Rev. Lett.}\ }\textbf {\bibinfo {volume} {104}},\ \bibinfo
  {pages} {150601} (\bibinfo {year} {2010})}\BibitemShut {NoStop}%
\bibitem [{\citenamefont {Tissier}\ and\ \citenamefont
  {Tarjus}(2011)}]{Tissier:2011zz}%
  \BibitemOpen
  \bibfield  {author} {\bibinfo {author} {\bibfnamefont {M.}~\bibnamefont
  {Tissier}}\ and\ \bibinfo {author} {\bibfnamefont {G.}~\bibnamefont
  {Tarjus}},\ }\href {\doibase 10.1103/PhysRevLett.107.041601} {\bibfield
  {journal} {\bibinfo  {journal} {Phys. Rev. Lett.}\ }\textbf {\bibinfo
  {volume} {107}},\ \bibinfo {pages} {041601} (\bibinfo {year} {2011})},\
  \Eprint {http://arxiv.org/abs/1103.4812} {arXiv:1103.4812
  [cond-mat.stat-mech]} \BibitemShut {NoStop}%
\bibitem [{\citenamefont {Benitez}\ \emph {et~al.}(2012)\citenamefont
  {Benitez}, \citenamefont {Blaizot}, \citenamefont {Chat\'e}, \citenamefont
  {Delamotte}, \citenamefont {M\'endez-Galain},\ and\ \citenamefont
  {Wschebor}}]{Benitez:2011xx}%
  \BibitemOpen
  \bibfield  {author} {\bibinfo {author} {\bibfnamefont {F.}~\bibnamefont
  {Benitez}}, \bibinfo {author} {\bibfnamefont {J.~P.}\ \bibnamefont
  {Blaizot}}, \bibinfo {author} {\bibfnamefont {H.}~\bibnamefont {Chat\'e}},
  \bibinfo {author} {\bibfnamefont {B.}~\bibnamefont {Delamotte}}, \bibinfo
  {author} {\bibfnamefont {R.}~\bibnamefont {M\'endez-Galain}}, \ and\ \bibinfo
  {author} {\bibfnamefont {N.}~\bibnamefont {Wschebor}},\ }\href {\doibase
  10.1103/PhysRevE.85.026707} {\bibfield  {journal} {\bibinfo  {journal} {Phys.
  Rev.}\ }\textbf {\bibinfo {volume} {E85}},\ \bibinfo {pages} {026707}
  (\bibinfo {year} {2012})},\ \Eprint {http://arxiv.org/abs/1110.2665}
  {arXiv:1110.2665 [cond-mat.stat-mech]} \BibitemShut {NoStop}%
\bibitem [{\citenamefont {Canet}\ \emph {et~al.}(2011)\citenamefont {Canet},
  \citenamefont {Chat\'e}, \citenamefont {Delamotte},\ and\ \citenamefont
  {Wschebor}}]{Canet11}%
  \BibitemOpen
  \bibfield  {author} {\bibinfo {author} {\bibfnamefont {L.}~\bibnamefont
  {Canet}}, \bibinfo {author} {\bibfnamefont {H.}~\bibnamefont {Chat\'e}},
  \bibinfo {author} {\bibfnamefont {B.}~\bibnamefont {Delamotte}}, \ and\
  \bibinfo {author} {\bibfnamefont {N.}~\bibnamefont {Wschebor}},\ }\href
  {\doibase 10.1103/PhysRevE.84.061128} {\bibfield  {journal} {\bibinfo
  {journal} {Phys. Rev. E}\ }\textbf {\bibinfo {volume} {84}},\ \bibinfo
  {pages} {061128} (\bibinfo {year} {2011})}\BibitemShut {NoStop}%
\bibitem [{\citenamefont {L\'eonard}\ and\ \citenamefont
  {Delamotte}(2015)}]{Leonard15}%
  \BibitemOpen
  \bibfield  {author} {\bibinfo {author} {\bibfnamefont {F.}~\bibnamefont
  {L\'eonard}}\ and\ \bibinfo {author} {\bibfnamefont {B.}~\bibnamefont
  {Delamotte}},\ }\href {\doibase 10.1103/PhysRevLett.115.200601} {\bibfield
  {journal} {\bibinfo  {journal} {Phys. Rev. Lett.}\ }\textbf {\bibinfo
  {volume} {115}},\ \bibinfo {pages} {200601} (\bibinfo {year}
  {2015})}\BibitemShut {NoStop}%
\bibitem [{\citenamefont {Rose}\ \emph {et~al.}(2016)\citenamefont {Rose},
  \citenamefont {Benitez}, \citenamefont {L\'eonard},\ and\ \citenamefont
  {Delamotte}}]{Rose:2016wqz}%
  \BibitemOpen
  \bibfield  {author} {\bibinfo {author} {\bibfnamefont {F.}~\bibnamefont
  {Rose}}, \bibinfo {author} {\bibfnamefont {F.}~\bibnamefont {Benitez}},
  \bibinfo {author} {\bibfnamefont {F.}~\bibnamefont {L\'eonard}}, \ and\
  \bibinfo {author} {\bibfnamefont {B.}~\bibnamefont {Delamotte}},\ }\href
  {\doibase 10.1103/PhysRevD.93.125018} {\bibfield  {journal} {\bibinfo
  {journal} {Phys. Rev.}\ }\textbf {\bibinfo {volume} {D93}},\ \bibinfo {pages}
  {125018} (\bibinfo {year} {2016})},\ \Eprint
  {http://arxiv.org/abs/1604.05285} {arXiv:1604.05285 [cond-mat.stat-mech]}
  \BibitemShut {NoStop}%
\bibitem [{\citenamefont {Balog}\ \emph {et~al.}(2019)\citenamefont {Balog},
  \citenamefont {Chat\'e}, \citenamefont {Delamotte}, \citenamefont
  {Marohnic},\ and\ \citenamefont {Wschebor}}]{Balog:2019rrg}%
  \BibitemOpen
  \bibfield  {author} {\bibinfo {author} {\bibfnamefont {I.}~\bibnamefont
  {Balog}}, \bibinfo {author} {\bibfnamefont {H.}~\bibnamefont {Chat\'e}},
  \bibinfo {author} {\bibfnamefont {B.}~\bibnamefont {Delamotte}}, \bibinfo
  {author} {\bibfnamefont {M.}~\bibnamefont {Marohnic}}, \ and\ \bibinfo
  {author} {\bibfnamefont {N.}~\bibnamefont {Wschebor}},\ }\href {\doibase
  10.1103/PhysRevLett.123.240604} {\bibfield  {journal} {\bibinfo  {journal}
  {Phys. Rev. Lett.}\ }\textbf {\bibinfo {volume} {123}},\ \bibinfo {pages}
  {240604} (\bibinfo {year} {2019})},\ \Eprint
  {http://arxiv.org/abs/1907.01829} {arXiv:1907.01829 [cond-mat.stat-mech]}
  \BibitemShut {NoStop}%
\bibitem [{\citenamefont {De~Polsi}\ \emph {et~al.}(2020)\citenamefont
  {De~Polsi}, \citenamefont {Balog}, \citenamefont {Tissier},\ and\
  \citenamefont {Wschebor}}]{DePolsi:2020pjk}%
  \BibitemOpen
  \bibfield  {author} {\bibinfo {author} {\bibfnamefont {G.}~\bibnamefont
  {De~Polsi}}, \bibinfo {author} {\bibfnamefont {I.}~\bibnamefont {Balog}},
  \bibinfo {author} {\bibfnamefont {M.}~\bibnamefont {Tissier}}, \ and\
  \bibinfo {author} {\bibfnamefont {N.}~\bibnamefont {Wschebor}},\ }\href@noop
  {} {\  (\bibinfo {year} {2020})},\ \Eprint {http://arxiv.org/abs/2001.07525}
  {arXiv:2001.07525 [cond-mat.stat-mech]} \BibitemShut {NoStop}%
\bibitem [{\citenamefont {Parola}\ and\ \citenamefont
  {Reatto}(1995)}]{Parola95}%
  \BibitemOpen
  \bibfield  {author} {\bibinfo {author} {\bibfnamefont {A.}~\bibnamefont
  {Parola}}\ and\ \bibinfo {author} {\bibfnamefont {L.}~\bibnamefont
  {Reatto}},\ }\href {\doibase 10.1080/00018739500101536} {\bibfield  {journal}
  {\bibinfo  {journal} {Advances in Physics}\ }\textbf {\bibinfo {volume}
  {44}},\ \bibinfo {pages} {211} (\bibinfo {year} {1995})},\ \Eprint
  {http://arxiv.org/abs/https://doi.org/10.1080/00018739500101536}
  {https://doi.org/10.1080/00018739500101536} \BibitemShut {NoStop}%
\bibitem [{\citenamefont {Canet}\ \emph
  {et~al.}(2004{\natexlab{a}})\citenamefont {Canet}, \citenamefont {Delamotte},
  \citenamefont {Deloubri\`ere},\ and\ \citenamefont
  {Wschebor}}]{Canet:2003yu}%
  \BibitemOpen
  \bibfield  {author} {\bibinfo {author} {\bibfnamefont {L.}~\bibnamefont
  {Canet}}, \bibinfo {author} {\bibfnamefont {B.}~\bibnamefont {Delamotte}},
  \bibinfo {author} {\bibfnamefont {O.}~\bibnamefont {Deloubri\`ere}}, \ and\
  \bibinfo {author} {\bibfnamefont {N.}~\bibnamefont {Wschebor}},\ }\href
  {\doibase 10.1103/PhysRevLett.92.195703} {\bibfield  {journal} {\bibinfo
  {journal} {Phys. Rev. Lett.}\ }\textbf {\bibinfo {volume} {92}},\ \bibinfo
  {pages} {195703} (\bibinfo {year} {2004}{\natexlab{a}})},\ \Eprint
  {http://arxiv.org/abs/cond-mat/0309504} {arXiv:cond-mat/0309504
  [cond-mat.stat-mech]} \BibitemShut {NoStop}%
\bibitem [{\citenamefont {Canet}\ \emph
  {et~al.}(2004{\natexlab{b}})\citenamefont {Canet}, \citenamefont {Chat\'e},\
  and\ \citenamefont {Delamotte}}]{Canet:2004je}%
  \BibitemOpen
  \bibfield  {author} {\bibinfo {author} {\bibfnamefont {L.}~\bibnamefont
  {Canet}}, \bibinfo {author} {\bibfnamefont {H.}~\bibnamefont {Chat\'e}}, \
  and\ \bibinfo {author} {\bibfnamefont {B.}~\bibnamefont {Delamotte}},\ }\href
  {\doibase 10.1103/PhysRevLett.92.255703} {\bibfield  {journal} {\bibinfo
  {journal} {Phys. Rev. Lett.}\ }\textbf {\bibinfo {volume} {92}},\ \bibinfo
  {pages} {255703} (\bibinfo {year} {2004}{\natexlab{b}})},\ \Eprint
  {http://arxiv.org/abs/cond-mat/0403423} {arXiv:cond-mat/0403423 [cond-mat]}
  \BibitemShut {NoStop}%
\bibitem [{\citenamefont {Caillol}(2006)}]{Caillol06}%
  \BibitemOpen
  \bibfield  {author} {\bibinfo {author} {\bibfnamefont {J.-M.}\ \bibnamefont
  {Caillol}},\ }\href {\doibase 10.1080/00268970600740774} {\bibfield
  {journal} {\bibinfo  {journal} {Molecular Physics}\ }\textbf {\bibinfo
  {volume} {104}},\ \bibinfo {pages} {1931} (\bibinfo {year} {2006})},\ \Eprint
  {http://arxiv.org/abs/https://doi.org/10.1080/00268970600740774}
  {https://doi.org/10.1080/00268970600740774} \BibitemShut {NoStop}%
\bibitem [{\citenamefont {Machado}\ and\ \citenamefont
  {Dupuis}(2010)}]{Machado:2010wi}%
  \BibitemOpen
  \bibfield  {author} {\bibinfo {author} {\bibfnamefont {T.}~\bibnamefont
  {Machado}}\ and\ \bibinfo {author} {\bibfnamefont {N.}~\bibnamefont
  {Dupuis}},\ }\href {\doibase 10.1103/PhysRevE.82.041128} {\bibfield
  {journal} {\bibinfo  {journal} {Phys. Rev.}\ }\textbf {\bibinfo {volume}
  {E82}},\ \bibinfo {pages} {041128} (\bibinfo {year} {2010})},\ \Eprint
  {http://arxiv.org/abs/1004.3651} {arXiv:1004.3651 [cond-mat.stat-mech]}
  \BibitemShut {NoStop}%
\bibitem [{\citenamefont {Ran\ifmmode~\mbox{\c{c}}\else \c{c}\fi{}on}\ and\
  \citenamefont {Dupuis}(2012)}]{Rancon12}%
  \BibitemOpen
  \bibfield  {author} {\bibinfo {author} {\bibfnamefont {A.}~\bibnamefont
  {Ran\ifmmode~\mbox{\c{c}}\else \c{c}\fi{}on}}\ and\ \bibinfo {author}
  {\bibfnamefont {N.}~\bibnamefont {Dupuis}},\ }\href {\doibase
  10.1103/PhysRevA.85.011602} {\bibfield  {journal} {\bibinfo  {journal} {Phys.
  Rev. A}\ }\textbf {\bibinfo {volume} {85}},\ \bibinfo {pages} {011602}
  (\bibinfo {year} {2012})}\BibitemShut {NoStop}%
\bibitem [{\citenamefont {Stevenson}(1981)}]{Stevenson:1981vj}%
  \BibitemOpen
  \bibfield  {author} {\bibinfo {author} {\bibfnamefont {P.~M.}\ \bibnamefont
  {Stevenson}},\ }\href {\doibase 10.1103/PhysRevD.23.2916} {\bibfield
  {journal} {\bibinfo  {journal} {Phys. Rev.}\ }\textbf {\bibinfo {volume}
  {D23}},\ \bibinfo {pages} {2916} (\bibinfo {year} {1981})}\BibitemShut
  {NoStop}%
\bibitem [{\citenamefont {Canet}\ \emph
  {et~al.}(2003{\natexlab{b}})\citenamefont {Canet}, \citenamefont {Delamotte},
  \citenamefont {Mouhanna},\ and\ \citenamefont {Vidal}}]{Canet:2002gs}%
  \BibitemOpen
  \bibfield  {author} {\bibinfo {author} {\bibfnamefont {L.}~\bibnamefont
  {Canet}}, \bibinfo {author} {\bibfnamefont {B.}~\bibnamefont {Delamotte}},
  \bibinfo {author} {\bibfnamefont {D.}~\bibnamefont {Mouhanna}}, \ and\
  \bibinfo {author} {\bibfnamefont {J.}~\bibnamefont {Vidal}},\ }\href
  {\doibase 10.1103/PhysRevD.67.065004} {\bibfield  {journal} {\bibinfo
  {journal} {Phys. Rev.}\ }\textbf {\bibinfo {volume} {D67}},\ \bibinfo {pages}
  {065004} (\bibinfo {year} {2003}{\natexlab{b}})},\ \Eprint
  {http://arxiv.org/abs/hep-th/0211055} {arXiv:hep-th/0211055 [hep-th]}
  \BibitemShut {NoStop}%
\bibitem [{\citenamefont {Guida}\ and\ \citenamefont
  {Zinn-Justin}(1998)}]{Guida:1998bx}%
  \BibitemOpen
  \bibfield  {author} {\bibinfo {author} {\bibfnamefont {R.}~\bibnamefont
  {Guida}}\ and\ \bibinfo {author} {\bibfnamefont {J.}~\bibnamefont
  {Zinn-Justin}},\ }\href {\doibase 10.1088/0305-4470/31/40/006} {\bibfield
  {journal} {\bibinfo  {journal} {J. Phys.}\ }\textbf {\bibinfo {volume}
  {A31}},\ \bibinfo {pages} {8103} (\bibinfo {year} {1998})},\ \Eprint
  {http://arxiv.org/abs/cond-mat/9803240} {arXiv:cond-mat/9803240 [cond-mat]}
  \BibitemShut {NoStop}%
\bibitem [{\citenamefont {Kompaniets}\ and\ \citenamefont
  {Panzer}(2017)}]{Kompaniets:2017yct}%
  \BibitemOpen
  \bibfield  {author} {\bibinfo {author} {\bibfnamefont {M.~V.}\ \bibnamefont
  {Kompaniets}}\ and\ \bibinfo {author} {\bibfnamefont {E.}~\bibnamefont
  {Panzer}},\ }\href {\doibase 10.1103/PhysRevD.96.036016} {\bibfield
  {journal} {\bibinfo  {journal} {Phys. Rev.}\ }\textbf {\bibinfo {volume}
  {D96}},\ \bibinfo {pages} {036016} (\bibinfo {year} {2017})},\ \Eprint
  {http://arxiv.org/abs/1705.06483} {arXiv:1705.06483 [hep-th]} \BibitemShut
  {NoStop}%
\bibitem [{\citenamefont {Delamotte}\ \emph {et~al.}(2016)\citenamefont
  {Delamotte}, \citenamefont {Tissier},\ and\ \citenamefont
  {Wschebor}}]{delamotte2016scale}%
  \BibitemOpen
  \bibfield  {author} {\bibinfo {author} {\bibfnamefont {B.}~\bibnamefont
  {Delamotte}}, \bibinfo {author} {\bibfnamefont {M.}~\bibnamefont {Tissier}},
  \ and\ \bibinfo {author} {\bibfnamefont {N.}~\bibnamefont {Wschebor}},\
  }\href@noop {} {\bibfield  {journal} {\bibinfo  {journal} {Physical Review
  E}\ }\textbf {\bibinfo {volume} {93}},\ \bibinfo {pages} {012144} (\bibinfo
  {year} {2016})}\BibitemShut {NoStop}%
\bibitem [{\citenamefont {Rosten}(2019)}]{Rosten:2016zap}%
  \BibitemOpen
  \bibfield  {author} {\bibinfo {author} {\bibfnamefont {O.~J.}\ \bibnamefont
  {Rosten}},\ }\href {\doibase 10.1142/S0217751X19500271} {\bibfield  {journal}
  {\bibinfo  {journal} {Int. J. Mod. Phys.}\ }\textbf {\bibinfo {volume}
  {A34}},\ \bibinfo {pages} {1950027} (\bibinfo {year} {2019})},\ \Eprint
  {http://arxiv.org/abs/1605.01729} {arXiv:1605.01729 [hep-th]} \BibitemShut
  {NoStop}%
\bibitem [{\citenamefont {De~Polsi}\ \emph {et~al.}(2019)\citenamefont
  {De~Polsi}, \citenamefont {Tissier},\ and\ \citenamefont
  {Wschebor}}]{DePolsi2019}%
  \BibitemOpen
  \bibfield  {author} {\bibinfo {author} {\bibfnamefont {G.}~\bibnamefont
  {De~Polsi}}, \bibinfo {author} {\bibfnamefont {M.}~\bibnamefont {Tissier}}, \
  and\ \bibinfo {author} {\bibfnamefont {N.}~\bibnamefont {Wschebor}},\ }\href
  {\doibase 10.1007/s10955-019-02411-3} {\bibfield  {journal} {\bibinfo
  {journal} {Journal of Statistical Physics}\ }\textbf {\bibinfo {volume}
  {177}},\ \bibinfo {pages} {1089} (\bibinfo {year} {2019})}\BibitemShut
  {NoStop}%
\bibitem [{Note2()}]{Note2}%
  \BibitemOpen
  \bibinfo {note} {By using Eq.~(\ref {eq_prop}), $H^{(n)}$ can be expressed in
  terms of propagators and vertex functions.}\BibitemShut {Stop}%
\bibitem [{Note3()}]{Note3}%
  \BibitemOpen
  \bibinfo {note} {There are 4 structures at order $\protect \mathcal O( p^5)$
  and 15 at order $\protect \mathcal O( p^7)$.}\BibitemShut {Stop}%
\bibitem [{Note4()}]{Note4}%
  \BibitemOpen
  \bibinfo {note} {We employed the equations of Refs.~\cite
  {delamotte2016scale,DePolsi2019} that differ with those of Ref.~\cite
  {Canet:2003qd} by terms of order $\protect \mathcal O(\partial
  ^6)$.}\BibitemShut {Stop}%
\bibitem [{\citenamefont {Kos}\ \emph {et~al.}(2014)\citenamefont {Kos},
  \citenamefont {Poland},\ and\ \citenamefont {Simmons-Duffin}}]{Kos:2014bka}%
  \BibitemOpen
  \bibfield  {author} {\bibinfo {author} {\bibfnamefont {F.}~\bibnamefont
  {Kos}}, \bibinfo {author} {\bibfnamefont {D.}~\bibnamefont {Poland}}, \ and\
  \bibinfo {author} {\bibfnamefont {D.}~\bibnamefont {Simmons-Duffin}},\ }\href
  {\doibase 10.1007/JHEP11(2014)109} {\bibfield  {journal} {\bibinfo  {journal}
  {JHEP}\ }\textbf {\bibinfo {volume} {11}},\ \bibinfo {pages} {109} (\bibinfo
  {year} {2014})},\ \Eprint {http://arxiv.org/abs/1406.4858} {arXiv:1406.4858
  [hep-th]} \BibitemShut {NoStop}%
\end{thebibliography}%

\end{document}